\documentclass[%
 reprint,
 amsmath,amssymb,
 aps,
]{revtex4-2}

\usepackage{graphicx}
\usepackage{dcolumn}
\usepackage{bm}
\usepackage[dvipsnames]{xcolor}

\begin{document}

\preprint{APS/123-QED}

\title{Granular focused jet}

\author{Kazuya U. Kobayashi}
\affiliation{%
 Department of Mechanical Engineering, Nippon Institute of Technology, 4-1 Gakuendai, Miyashiro-machi, Minamisaitama-gun, Saitama, Japan
}%

\author{Pradipto}
\author{Yoshiyuki Tagawa}%
\affiliation{
 Department of Mechanical Systems Engineering, Tokyo University of Agriculture and Technology, 2-24-16 Nakacho, Koganei-city, Tokyo, Japan
}%


\date{\today}

\begin{abstract}
We investigated the formation of granular focused jets defined as narrow upward ejections of grains triggered by impulsive forces and shaped by kinematic focusing on a concave free surface.
The jets were generated from nonfluidized granular beds, in contrast to existing granular Worthington jets that originate from fluidized layers.
To elucidate the jet dynamics, we performed experiments in which a test tube partially filled with dry glass beads was dropped onto a flat rigid floor, systematically varying the granular pile height$L_{\rm G}$, drop height $H$, and particle size.
The resulting jet formation in granular media was driven by the same kinematic focusing mechanism responsible for jetting in liquids.
By conducting parallel experiments using low-viscosity silicone oil under identical conditions, we directly compared the granular and liquid jets.
At low pile heights, the granular jet velocity quantitatively agreed with the liquid jet velocity.
However, at high pile heights, contrasting trends emerged. Specifically, the granular jet velocity decreased with increasing $L_{\rm G}$, while the liquid jet velocity increased due to cavitation.
Discrete element method simulations confirmed that the velocity reduction in granular jets arose from energy dissipation via grain--grain contacts during impact force propagation.
These findings highlight common mechanisms and distinctive dissipation behaviors in granular and liquid focused jets.
\end{abstract}

\maketitle

\section{Introduction}
Granular materials, which are defined as collections of solid particles of various sizes or aggregates of solid particles, are used in diverse fields such as civil engineering, chemical engineering, and earth sciences.
These materials are also found in natural phenomena such as avalanches, landslides, and earthquake-induced liquefaction. Moreover, they play a crucial role in industrial products and processes including pharmaceuticals, paints, and cosmetics.
Consequently, extensive interdisciplinary research has been conducted on granular materials \cite{Iverson1997, Duran1999, Coussot2005}.

The mechanical behavior of granular materials differs fundamentally from that of liquids.
For instance, granular materials dissipate energy through interparticle friction and inelastic collisions, lack surface tension, and contain particles whose sizes are often comparable to the characteristic flow lengths.
These features hinder the description of granular flows using continuum models based on fluid mechanics \cite{Jaeger1992, Herrmann1998, Gennes1999, Andreotti2013, Pradipto2021, Kobayashi2022}.
In addition, forces in granular media are transmitted through networks of contacting particles, forming chain-like structures known as force chains.
These force chains lead to strongly inhomogeneous stress distributions, and the overall mechanical response is sensitive to particle packing and deformation history \cite{Jaeger1996, Duran1999, Trushant2005}.

Despite these differences, granular materials exhibit various flow behaviors that resemble those of fluids.
A well-known example is the formation of granular jets when a solid object impacts a loosely packed granular bed.
The impact generates a broad splash, followed by the formation of a transient axisymmetric crater and finally a narrow upward jet of particles \cite{Thoroddsen2001, Lohse2004, Royer2005, Royer2008, Katsuragi2010, Devaraj2017}.
This phenomenon is analogous to Worthington jets observed in liquids \cite{Gekle2010, Devaraj2017}.
These so-called granular Worthington jets are typically observed in fluidized granular beds, where gas injection reduces the packing fraction to approximately 0.4--0.5 \cite{Lohse2004, Royer2005, Royer2008}.
However, in densely packed (nonfluidized) beds with higher packing fractions, the solid object cannot penetrate deeply enough to form a cavity, and jet formation is generally suppressed.

Recent studies have shown the possibility of generating narrow focused jets from nonfluidized granular materials if an impulsive force is applied to the bed surface \cite{Antkowiak2007, Sabuwala2018}.
These jets exhibit sharply elongated tips and are often called focused jets \cite{Eggers2008, Tagawa2012, Gordillo2020}.
The formation of such focused jets is not limited to granular materials, and it has been extensively studied in liquid systems.
For example, Pokrovski's experiments demonstrated that when a glass test tube filled with liquid falls freely and impacts a rigid surface, the interface near the liquid surface is sharply accelerated.
Owing to the concave shape formed by the liquid--gas interface during free fall, resulting from wetting effects on the tube wall, this impulsive acceleration produces a highly focused liquid jet.
This experimental setup is common for studying impulsive jet generation in liquids and serves as a reference for comparison with granular jet behavior.

Several studies have contributed to understanding the mechanisms underlying impulsive jet formation.
\citet{Antkowiak2007} solved the Laplace equation for an incompressible fluid subjected to an impulsive acceleration, successfully predicting the early time velocity field near the free surface.
Additionally, \citet{Kiyama2016} showed that increasing the liquid height or impact strength can trigger cavitation in the bulk, substantially enhancing the jet velocity.
These results emphasize the coupled effects of boundary geometry, fluid properties, and internal pressure dynamics on jet formation.

In granular materials, energy is dissipated primarily through inelastic collisions and friction, and compressibility arises from the rearrangement of discrete particles.
Although previous studies have reported the formation of granular jets under impulsive loading \cite{Antkowiak2007, Sabuwala2018}, the specific roles of compressibility and dissipation in determining the jet velocity and structure have yet to be fully clarified.

In this study, we investigated the formation of focused granular jets generated by impulsive forces, combining laboratory experiments with discrete element method (DEM) simulations.
We focused on initial granular pile height $L_{\rm G}$ as the main control parameter because it determines the extent of force propagation and energy dissipation within the medium.
This approach was motivated by previous studies on liquid jets, in which increasing the liquid column height led to unexpected jet velocity enhancement due to cavitation effects \cite{Kiyama2016}.
By systematically comparing granular and liquid jets under similar initial and boundary conditions, we aimed to reveal how dissipation and geometrical focusing interact to determine jet formation.
Through this comparison, we sought to identify the physical mechanisms unique to granular jets and clarify conditions under which their behavior converged or diverged from that of liquid jets.

\section{Methods}
\subsection{\label{sec:level2}Experimental Method}

\begin{figure}[htbp]
\begin{center}
\includegraphics[width=60mm]{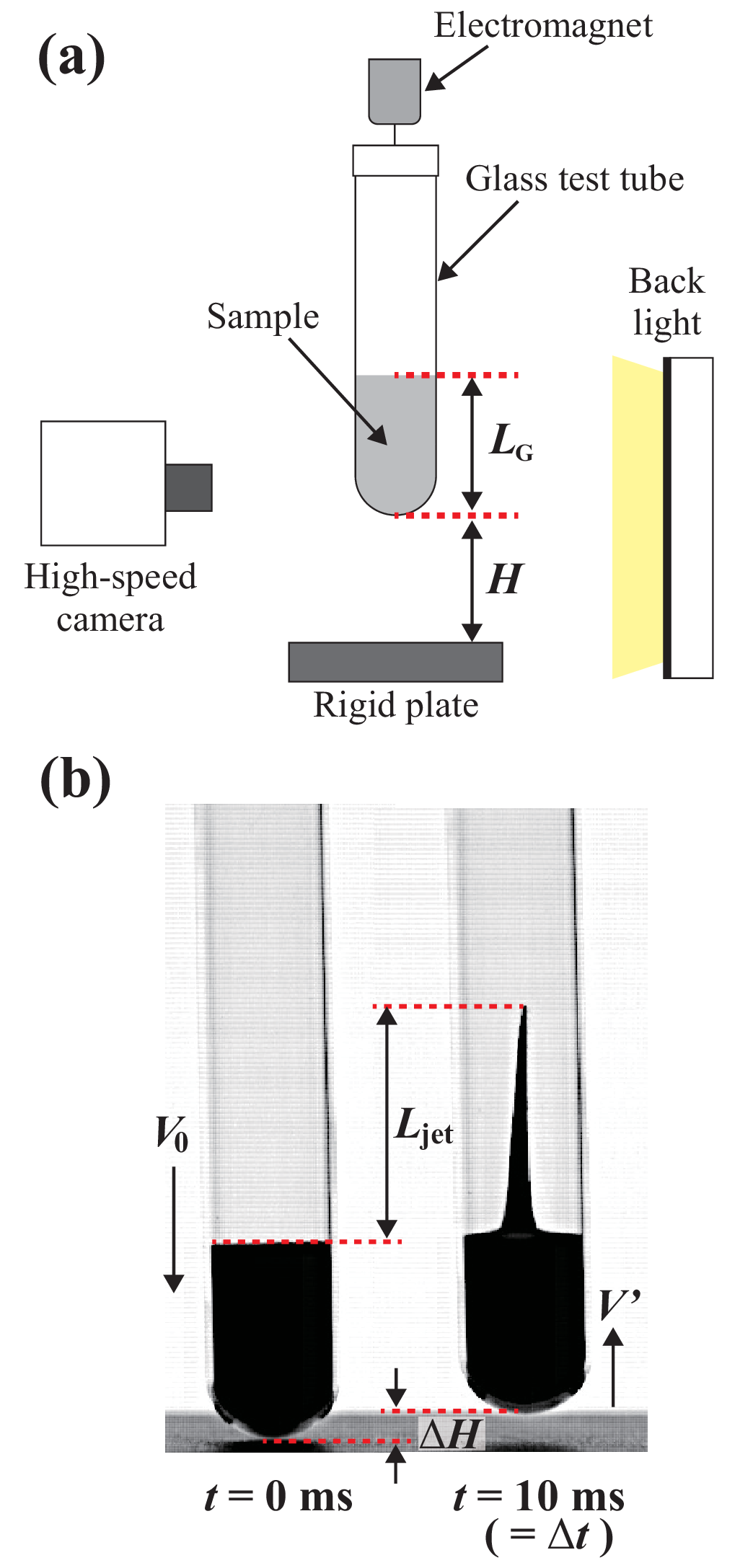}
\end{center}
\caption{\label{fig:1}
(Color online) (a) Schematic of experimental setup. A test tube containing liquid or granular material is dropped from height $H$ onto a metallic floor. The internal height of the sample is denoted as $L_{\rm G}$. (b) Image sequence showing jet formation and definitions of $V_0$, $V'$, $L_{\rm jet}$, and $\Delta H$, used to determine $U_0$ and $V_{\rm jet}$.}
\end{figure}

In experiments, we investigated jet formation by dropping a glass test tube filled with either silicone oil or dry glass beads from height $H$ onto a rigid floor, as illustrated in Fig.~\ref{fig:1}(a).
Upon impact, a jet emerged from the fluid or granular surface, and jet velocity $V_{\rm jet}$ was analyzed according to various experimental parameters.
The key physical quantities were pile height $L_{\rm G}$ (initial height of the fluid or grains), drop height $H$, and impact velocity $U_0$.

Impact velocity $U_0$ imparted to the system by the impulsive force and velocity $V_{\rm jet}$ of the generated jet are defined as follows \cite{Kiyama2016}:
\begin{eqnarray}
U_0=V_0+V'=\sqrt{2gH}+(\Delta H / \Delta t),
\label{eq:1}
\end{eqnarray}
\begin{eqnarray}
V_{\rm{jet}}=(L_{\rm{jet}}-\Delta H) / \Delta t,
\label{eq:2}
\end{eqnarray}
where $V_0$ is the velocity upon collision with the metallic floor surface, $V'$ is the velocity when the test tube rebounds after collision, $g$ is the gravitational acceleration, $H$ is the falling height of the test tube, $\Delta t$ is the elapsed time after impact (10~ms), $L_{\rm{jet}}$ is the distance from the initial interface to the tip of the generated jet at $\Delta t$, and $\Delta H$ is the rebound height from the metallic floor material at $\Delta t$.
These variables are depicted in Fig.~\ref{fig:1}(b).

A glass test tube was filled with glass beads or silicone oil at heights $L_{\rm G}$ from 25 to 100~mm.
The falling height $H$ of the test tube was varied from 10 to 150~mm.
The experiments were repeated five times under the same conditions, and the mean and standard deviation were considered as the representative value and error range, respectively.
The room temperature and humidity were maintained at approximately 20~$^{\circ}$C and 70~\%, respectively.

The test tube was made of borosilicate glass with an inner diameter of 14.2~mm and thickness of 1.2~mm.
A metallic hook was attached to a silicone cap mounted on the test tube. The test tube was suspended in the test apparatus using an electromagnet, and the axis of the tube was aligned with the vertical axis.
By turning off the electromagnet switch, the test tube fell freely from height $H$, i.e., distance from the bottom of the tube to the floor).
Then, focused jets were generated upon collision with a metallic floor made of SS400 steel.
Jet formation was recorded using a high-speed camera (Photron Co., Ltd., FASTCAM Nova R2).

Care was taken to ensure that the metallic hook attached to the silicone cap and electromagnet did not misalign, as this could prevent the test tube from tilting during free fall as the drop height increased.
In addition, to ensure consistent drop conditions, the experiment began after confirming that the test tube was stationary.
An evaluation of the test tube tilt during free fall demonstrated that the tilt was less than 5 degrees, and the variation in the jet velocity was confirmed to be less than 10~\% under the same preset experimental conditions, confirming the reproducibility of the experimental setup.

The granular material used in this experiment was spherical soda-lime glass beads (Fuji Manufacturing FGB-320 and FGB-240).
Based on the catalog values, two types of glass beads were selected for the experiment, one with a particle size range of 38--53~$\mu$m (denoted by its average size of 45~$\mu$m) and the other with a range of 63--75~$\mu$m (denoted by its average size of 70~$\mu$m).
Both types of beads had a true density of 2.5~g/cm$^3$ given by the mass of a particle per unit volume, excluding the volume of both open and closed pores.

According to the Geldart diagram \cite{Geldard1973, Cocco2023}, both types of particles belonged to group A, indicating that they were easily fluidized.
Similar to liquid systems, the glass beads were filled into the test tube with sample height $L_{\rm G}$ ranging from 25~mm to 100~mm.
In addition, to ensure that initial packing fraction $\phi$ of the granular material was consistent at the start of each experiment, the height of the material in the test tube was adjusted by tapping the tube several times.
We estimated the packing fraction as $\phi = V_{\rm g} / V_{\rm t}$, where $V_{\rm g}$ is the volume occupied by particles and $V_{\rm t}$ is the total volume, obtaining approximately 0.55.

During the tube free fall, for a perfectly wetting liquid, the liquid--gas interface quickly forms a hemisphere after the initiation of the free fall. However, this phenomenon is not observed in granular systems.
Unlike liquids, granular particles lack the wetting behavior on the inner surface of the test tube.
Instead, they possess effective surface tension due to van der Waals interactions, which allow for the preservation of the meniscus.
To replicate this behavior, a glass rod with a hemispherical shape matching the bottom of the test tube (diameter of 15~mm) was utilized to pre-shape the granular surface concavely.

As a liquid sample, we used silicone oil (KF-96L-10cs, density $\rho_{\rm s}$ of 0.82~g/cm$^3$, kinematic viscosity of 10~mm$^2$/s) manufactured by Shin-Etsu Chemical.
The silicone oil was degassed before the experiment to prevent the introduction of air bubbles.

\subsection{\label{subsec:simulation_method} Discrete Element Method Simulation}

We performed DEM simulations to elucidate the mechanism of the impact-induced granular jet until the microscopic level.
The DEM was solved using the open-source large-scale atomic/molecular massively parallel simulator \cite{lammps} improved for general granular and granular heat transfer simulations (LIGGGHTS) \cite{liggghts}.
The simulation involved spherical particles with diameter $D$, mass $m$, and density $\rho$, a cylindrical test tube with height $L_{\rm{T}}$ and diameter $D_{\rm{T}}$, and a hemispherical cap.
We ignored the particle--air interactions inside the test tube. However, for particles with diameter less than 100~$\mu m$, the van der Waals cohesive forces between grains could not be ignored \cite{partelli2014}.
Thus, we implemented the Johnson--Kendall--Roberts (JKR) model \cite{johnson1971} using the JKR patch in LIGGGHTS \cite{eidevag2019}.
In addition, we considered gravity, sliding friction with coefficient $\mu$, and rolling friction $\mu_r$ for the particle--particle interactions.
We calibrated the simulation parameters using the angle of repose test.
LIGGGHTS triangulated the surface of the tube and cap into meshes. In particular, it discretized the walls (e.g., tube and cap walls) into triangular mesh elements and resolved particle--wall interactions by independently solving the interaction between particles and each mesh element.
For particle--mesh interactions, we adopted the Hertzian contact with sliding friction and rolling friction with the same $\mu$ and $\mu_r$ settings as in the particle--particle interactions.

\begin{table}[htbp]
    \centering
    \begin{tabular}{c c c}
        Parameter & Value \\
            \hline
         Density $\rho$ & 2500 $\rm kg / m^3$  \\
         Young's modulus $E$ & 630 MPa \\
         Restitution coefficient $\epsilon$ & 0.2\\
         Poisson's ratio $\nu$ & 0.24 \\
         Friction coefficient $\mu$ & 0.5\\
         Rolling friction coefficient $\mu_r$ & 0.05\\
         JKR surface tension  $\gamma$ & $1\times10^{-3}$ $\rm N/m^2$\\
    \end{tabular}
    \caption{Simulation parameters.}
    \label{tab:1}
\end{table}

The simulation protocol utilized in this study is described as follows.
First, we filled the tube with $N=20,000$ particles, resulting in a granular pile with height $L_{\rm{G}}$.
Packing fraction of the granular pile, $\phi \approx 0.574$, was adopted for all simulations.
We then created an artificial meniscus by pushing the bed with a hemispherical cap until initial contact angle $\theta \approx 30^{\circ}$ was reached.
After the energy of the system relaxed, we pulled the cap and allowed the surface to stabilize.
Then, we allowed the system to fall freely from height $H_0$ before the sudden stop, with $H$ corresponding to impact velocity $U_0 = \sqrt{2gH}$ for gravitational acceleration $g$.
The interaction between the tube and floor was not considered in these simulations. Thus, the only driving force was the inertial force due to the sudden stop after the free fall.

\begin{figure*}[htbp]
\begin{center}
\includegraphics[width=155mm]{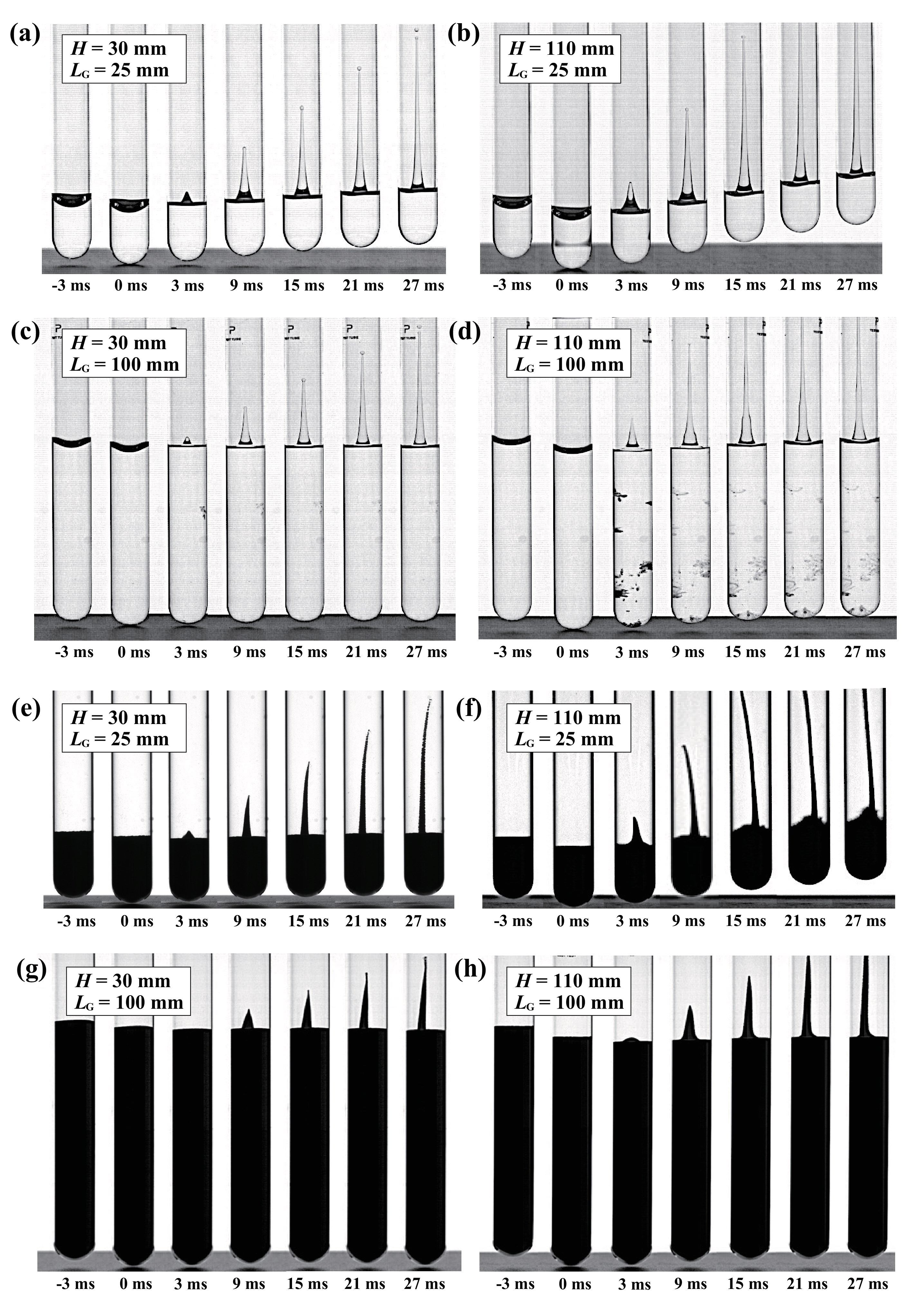}
\end{center}
\vspace{-5mm}
\caption{\label{fig:2}
Time evolution of jet formation in (a)--(d) liquid and (e)--(h) granular systems (particle size of 45~$\mu$m). 
Each column shows snapshots at $t$ = -3, 0, 3, 9, 15, 21, and 27~ms, with impact occurring at $t$ = 0~ms. 
The conditions include two cases with fixed $L_{\rm G}$ (25 and 100~mm) and varying $H$, and two with fixed $H$ (30 and 110~mm) and varying $L_{\rm G}$, enabling direct comparisons between liquid and granular jets.
At $L_{\rm G}$ = 25~mm, jet velocities are similar for both $H$ values.
For $H$ = 110~mm, the liquid jet velocity increases with $L_{\rm G}$ due to cavitation, while the granular jet velocity decreases, reflecting different underlying mechanisms.}
\end{figure*}

\begin{figure*}[htbp]
\begin{center}
\includegraphics[width=130mm]{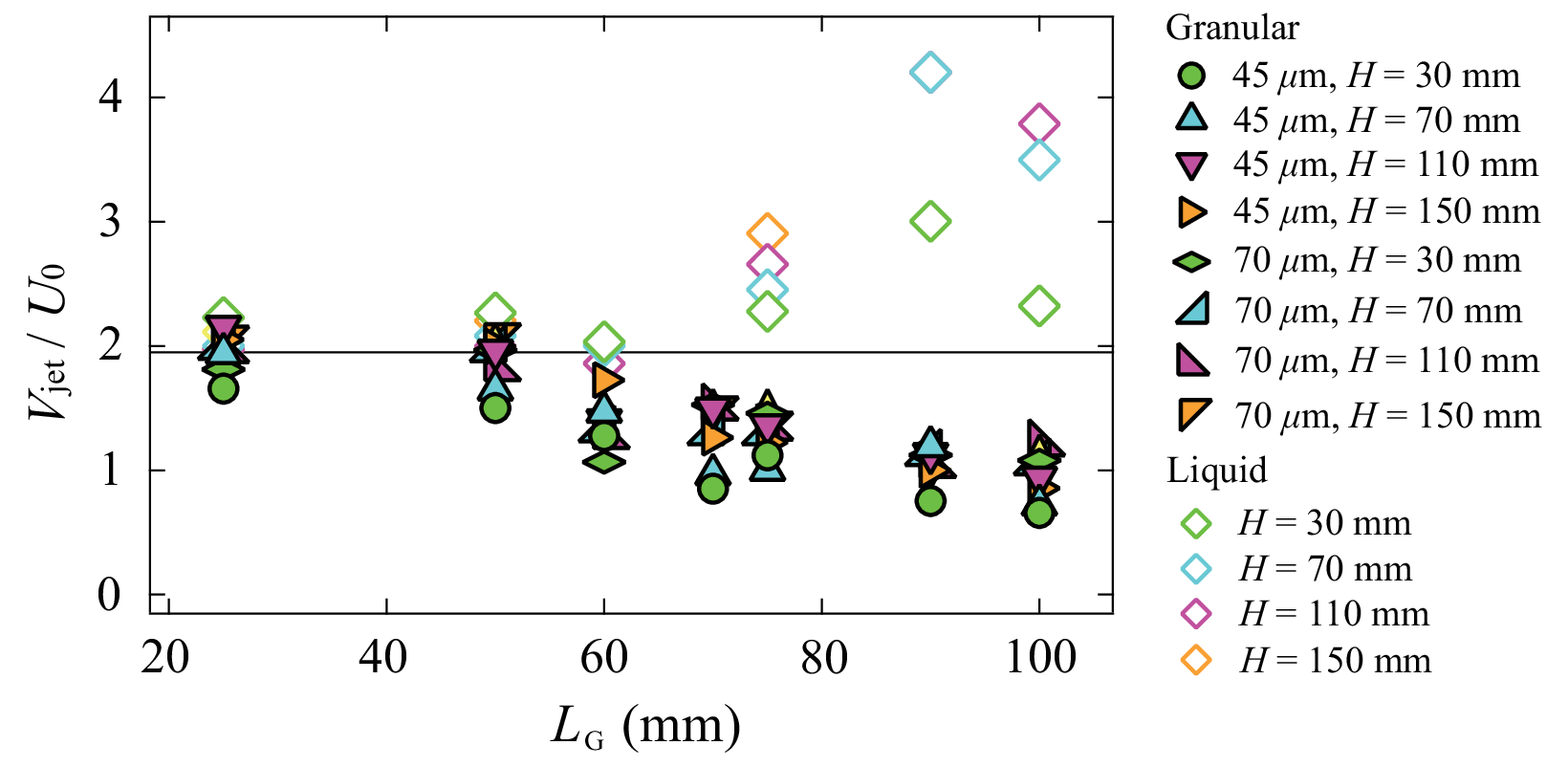}
\end{center}
\caption{\label{fig:3}
(Color online) Normalized jet velocity $V_{\rm{jet}}/U_0$ according to granular/liquid pile height $L_{\rm G}$ for various drop heights $H$.
Filled markers represent granular jets, and open markers represent liquid jets.
The black curve indicates $V_{\rm{jet}}/U_0 \approx 1.95 \sim 2$, as reported in \cite{Tagawa2012, Kiyama2016, Kurihara2025}}.
\end{figure*}

\section{Results and Discussions}

\subsection{\label{Experimental Results}Experimental Results}
\subsubsection{Jet behavior in liquid and granular systems}

We present representative experimental results for both the liquid and granular materials.
In Fig.~\ref{fig:2}, $t = 0$~ms is defined as the moment of collision between the test tube and floor.

Figures~\ref{fig:2}(a)--(d) show jet formation in silicone oil under four representative conditions, namely, two cases with fixed $L_{\rm G}$ and varying $H$, and two cases with fixed $H$ and varying $L_{\rm G}$. These conditions enable direct comparison with granular jet experiments.
During the tube free fall, the liquid--gas interface deforms into a hemispherical shape ($t$ = -3~ms).
Upon collision with the metallic floor, the interface rapidly inverts, producing a focused liquid jet ($t$ = 3~ms).
This focused jet elongates and gradually breaks into droplets ($t$ = 9--27~ms).
In liquid jets, an impulsive force rapidly accelerates the liquid. Near a concave gas--liquid interface, flow is directed toward the axis in a process known as kinematic focusing, resulting in the formation of a liquid jet, whose diameter is narrower than that of the nozzle \cite{Antkowiak2007, Kiyama2016, Onuki2018}.
As the jet extends, it undergoes pinch-off into smaller droplets due to Rayleigh--Plateau instability (Fig.~\ref{fig:2}(a), $t$ = 21--27~ms).

The effect of $H$ and $L_{\rm G}$ on jet velocity reveals distinct behavioral regimes in the liquid system.
For $L_{\rm G}$ = 25~mm (Figs.~\ref{fig:2}(a) and (b)), the jet velocity clearly increases with falling height $H$.
For $L_{\rm G}$ = 100~mm (Figs.~\ref{fig:2}(c) and (d)), increasing $H$ not only enhances the jet velocity but also leads to the formation of cavitation bubbles within the liquid, further increasing the jet velocity.
When $H$ = 30~mm (Figs.~\ref{fig:2}(a) and (c)), the jet velocity remains nearly constant even under a large variation in $L_{\rm G}$.
In contrast, at larger $H$ (=110~mm, Figs.~\ref{fig:2}(b) and (d)), cavitation effects contribute substantially to increasing the jet velocity.
These trends are consistent with previous studies \cite{Antkowiak2007, Kiyama2016}.
Thus, in the liquid system, the jet velocity is generally independent of $L_{\rm G}$ at low $H$ but increases substantially with $H$.
Moreover, at high $L_{\rm G}$, cavitation further enhances the jet velocity.

Figures~\ref{fig:2}(e)--(h) show the granular jet behaviors under the same four conditions considered for the liquid.
At $t$ = -3~ms, no visible deformation occurs on the free surface.
After impact ($t$ = 3~ms), a jet emerges from the center of the test tube.
The granular jet elongates and breaks into clusters ($t$ = 9--27~ms), resembling the liquid case in shape and dynamics.
The granular packing fraction is $\phi$ = 0.55, characteristic of a nonfluidized state.

The jet formation behavior in the granular system also exhibits focused jet structures similar to those in the liquid system, particularly when the granular surface is pre-shaped as a hemisphere.
When a flat surface is used in granular systems (as opposed to a concave one), the surface particles scatter upward as a bulk rather than forming a focused jet, indicating the necessity of initial surface shaping.
For $L_{\rm G} = 25$~mm and low falling height $H$, a conical jet emerges after impact and subsequently elongates and fragments into clusters ($t$ = 9--27~ms), as shown in Fig.~\ref{fig:2}(e).
This behavior closely resembles that of the liquid jet shown in Fig.~\ref{fig:2}(a), both in jet shape and elongation dynamics.
Given packing fraction $\phi = 0.55$, which corresponds to a nonfluidized state, this result suggests that the pre-imposed surface curvature induces kinematic focusing even in dry granular media, leading to the formation of a focused jet tip.

Granular jets may exhibit lateral bending, likely caused by slight asymmetries in the initial surface or particle distribution.
However, initial ejection is vertical, and subsequent bending has a negligible impact on the jet axial velocity.
To ensure such ejection, we analyze jet velocities under the same $L_{\rm G}$ and $H$ for cases with lateral bending and confirm a variation within 10~\%.
Therefore, we conclude that the bending behavior does not notably influence the present discussion.

For fixed $L_{\rm G} = 25$~mm and increasing $H$, the granular jet velocity becomes noticeably higher (Fig.\ref{fig:2}(f)), resembling the velocity enhancement observed in liquids (Fig.\ref{fig:2}(b)).
This finding confirms that falling height $H$ governs the strength of inertial focusing in both systems despite the absence of surface tension in the granular medium.

A different trend appears when the granular pile height increases.
At $H = 30$~mm and $L_{\rm G} = 100$~mm, a focused jet still forms (Fig.\ref{fig:2}(g)), but the resulting velocity is substantially reduced compared with the case at $L_{\rm G} = 25$~mm.
This result contrasts sharply with the case of liquids, in which the jet velocity is relatively insensitive to variations in $L_{\rm G}$ under low $H$ (Fig.~\ref{fig:2}(c)).

Furthermore, under high $H$, the granular jet velocity increases slightly with $H$ (Fig.\ref{fig:2}(h)), but it remains substantially lower than the corresponding liquid jet velocity (Fig.\ref{fig:2}(d)).
This difference can be attributed to the absence of cavitation in the granular medium, which contributes to the dramatic velocity enhancement in liquid systems.
Thus, the granular system exhibits a distinct behavior. The jet velocity decreases with increasing $L_{\rm G}$ and slightly increases with $H$, lacking the cavitation-driven amplification observed in liquids.

To clarify these contrasting behaviors between the two systems, normalized jet velocity $V_{\rm jet} / U_0$ is analyzed in the following section.

\begin{figure*}[htbp]
\begin{center}
\includegraphics[width=0.9\linewidth]{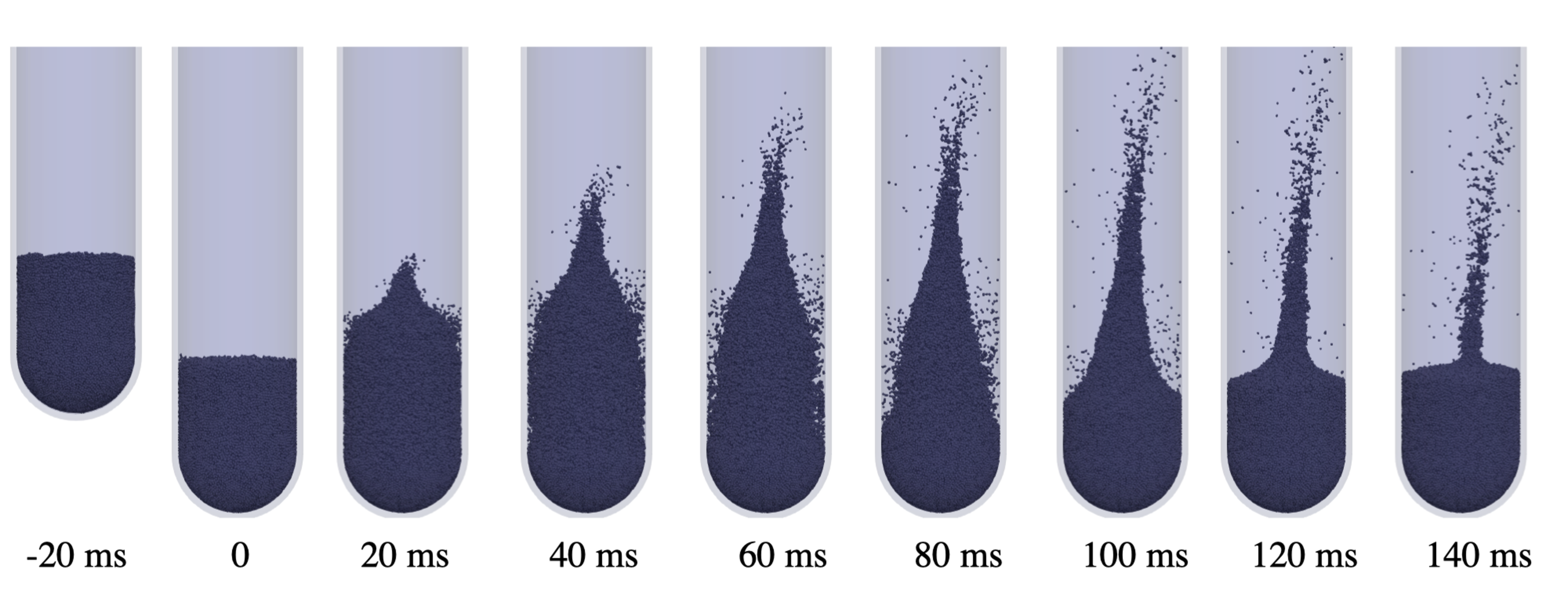}
\end{center}
\caption{\label{fig:4}
Time evolution of impact-induced granular jet obtained from DEM simulation.
An initial concave meniscus is imposed on the granular surface to replicate the experimental conditions.
After impact, a focused granular jet forms, exhibiting slight lateral bending during ascent, consistent with experimental observations.}
\end{figure*}

\subsubsection{Dependence of jet velocity on granular pile height $L_{\rm G}$}

We evaluated the variation in normalized jet velocity $V_{\rm jet} / U_0$ according to granular pile height $L_{\rm G}$, falling height $H$, and the particle size for both liquid and granular systems.
The results are summarized in Fig.~\ref{fig:3}, where filled markers correspond to granular jets, while open markers correspond to liquid jets.

Overall, the normalized jet velocity of the granular system remains nearly constant at $V_{\rm jet}/U_0 \approx 2$ for $L_{\rm G} \leq 50$~mm, similar to liquid jets without cavitation.
However, for $L_{\rm G} > 50$~mm, the granular jet velocity drastically decreases, while the normalized velocity of the liquid jet continues to increase.

This initial plateau for $L_{\rm G} \leq 50$~mm suggests that at shallow granular depths, the inertial energy imparted by the impulsive force is efficiently focused into jet formation, similar to kinematic focusing in liquids.
In this regime, the particle size and internal dissipation appear to have a negligible impact, and the granular system behaves similarly to an ideal inviscid liquid.

Beyond $L_{\rm G}$ = 50 mm, $V_{\rm jet} / U_0$ for particles of both 45~$\mu$m and 70~$\mu$m begins to decline, with a sharper decrease observed for smaller particles.
This trend highlights the growing importance of interparticle interactions such as friction and inelastic collisions, which become increasingly prominent as the number of grain--grain contact points grows with $L_{\rm G}$.
Furthermore, the lower velocities observed for 45~$\mu$m particles compared with 70~$\mu$m particles suggest that smaller particles are more susceptible to energy dissipation, possibly due to the larger contact surface area and total contact frequency.

These observations indicate that both the particle size and granular layer thickness $L_{\rm G}$ affect the internal dissipation mechanisms, which in turn govern jet formation.

In the liquid system, the jet velocity follows linear relation $V_{\rm jet} = \alpha U_0$, being consistent with potential flow theory (or the pressure impulse approach), with $\alpha \approx 1.95$ for $20 < L_{\rm G} \leq 60$~mm, as shown in Fig.~\ref{fig:3}.

For $L_{\rm G} > 60$~mm, the liquid system exhibits a considerable increase in jet velocity primarily due to the cavitation onset.
Previous studies have shown that cavitation bubbles in bulk liquid can substantially accelerate a jet, although the resulting velocity may fluctuate depending on the maximum size and dynamics of these bubbles\cite{Kiyama2016}.
Velocity amplification due to cavitation in the liquid system contrasts with the behavior in the granular system, which shows a monotonic decrease in jet velocity with increasing $L_{\rm G}$.

The difference in behavior between the two systems can be partially explained by the cavitation number, $Ca$, which predicts the likelihood of cavitation in accelerating liquids \cite{Pan2017}.
The cavitation number is defined as
\begin{eqnarray}
Ca = \frac{p_r - p_v}{\rho a L_{\rm G}},
\end{eqnarray}
where $p_r$ is the ambient pressure, $p_v$ is the vapor pressure of the liquid, $\rho$ is the liquid density, $a$ is the acceleration of the liquid, and $L_{\rm G}$ is the height of the liquid column. Cavitation is expected to occur when $Ca < 1$.

In our experiments, $L_{\rm G} \gtrsim 50$~mm corresponds to the regime where $Ca < 1$, making cavitation highly probable.
This justifies the observed velocity increase in the liquid jets for large $L_{\rm G}$ values and explains the dominance of cavitation in the dynamics.

In contrast, cavitation does not occur in granular materials.
Instead, the decrease in $V_{\rm jet} / U_0$ with increasing $L_{\rm G}$ is attributed to increased energy dissipation caused by intensified grain--grain interactions.
As the granular layer thickens, the number of particle contacts increases, thereby enhancing the internal friction and collisional loss, which in turn reduce the available energy for jet formation.
This mechanism has been reported in \cite{Essayah2022, Zhao2008} and motivates the use of DEM simulations to further investigate dissipation effects during granular jet formation, as discussed in the next section.

\begin{figure}[htbp]
\begin{center}
\includegraphics[width=0.9\linewidth]{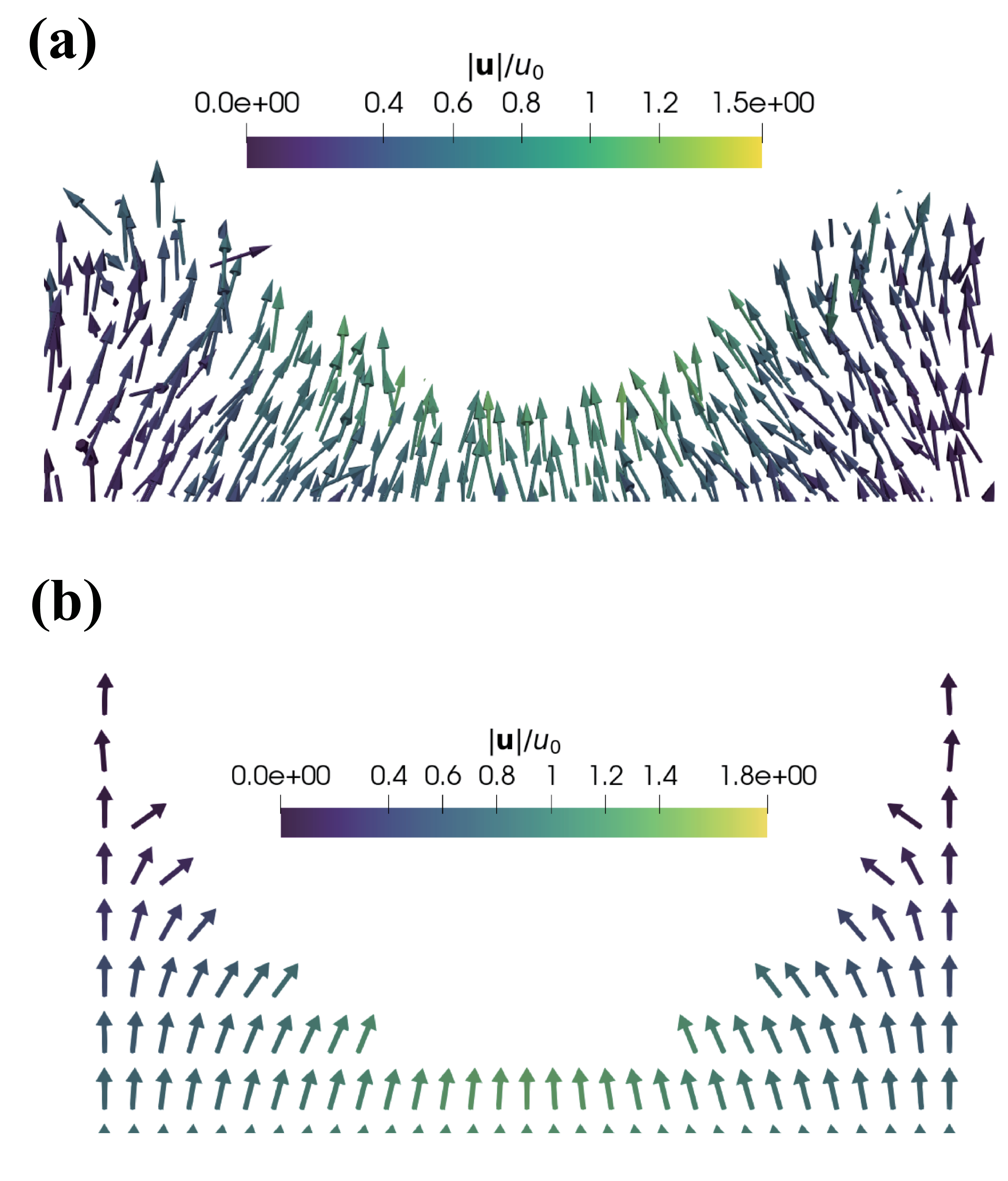}
\end{center}
\caption{\label{fig:5} (Color online) Snapshot of velocity field near meniscus immediately after impact. (a) Granular jet obtained from DEM simulation and (b) liquid jet calculated using potential flow theory.
The granular velocity field qualitatively resembles the potential flow solution, indicating the occurrence of kinematic focusing in granular jets.}
\end{figure}

\subsection{\label{DEM Simulation Results}Simulation Results}
\subsubsection{Jet formation and dependence on pile height}

The evolution of the impact-induced granular jet in the DEM simulation is shown in Fig.~\ref{fig:4}, demonstrating a behavior qualitatively consistent with experimental observations.
Jet formation occurs only when a concave meniscus is imposed on the surface of the granular bed, highlighting the importance of the initial surface curvature in promoting the ejection of focused jets.

To further understand the internal flow structure, we analyze the velocity field immediately after impact.
Figure~\ref{fig:5}(a) shows a snapshot of the particle velocity vectors on a monolayer slice on the $x$--$z$ plane obtained by slicing the system along the center of the $y$-- axis.
The velocity pattern resembles the theoretical potential flow solution computed via the COMSOL software (Fig.~\ref{fig:5}(b)), confirming that kinematic focusing similar to that in fluids also occurs in the granular system.

\begin{figure}[htbp]
\begin{center}
\includegraphics[width=0.9\linewidth]{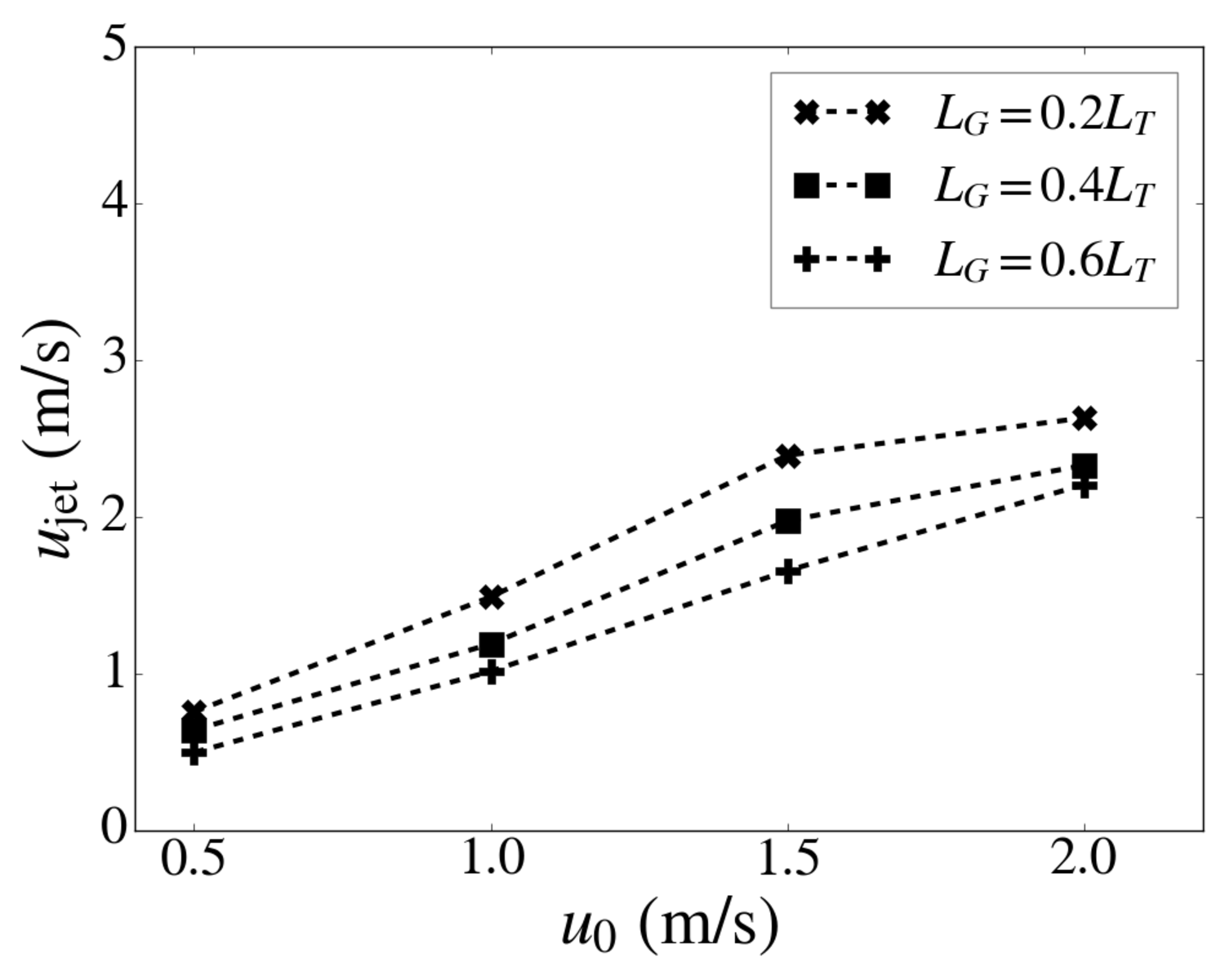}
\end{center}
\caption{\label{fig:6}
Relations between jet and impact velocities for various granular pile heights $L_{\rm G}$.
The jet velocity decreases with increasing $L_{\rm G}$, consistent with experimental observations.}
\end{figure}

In addition, the simulation results reproduce the dependence of the jet behavior on granular pile height $L_{\rm G}$, consistent with the experimental findings (see Fig.\ref{fig:2}).
By varying the initial pile height as $L_{\rm G} = 0.2L_{\rm T}$, $0.4L_{\rm T}$, and $0.6L_{\rm T}$---corresponding to $N=10,000$, $N=20,000$, and $N=35,000$ particles, respectively---we observed systematic changes in the jet dynamics.
As shown in Fig.~\ref{fig:6}, the jet velocity decreases as $L_{\rm G}$ increases, reinforcing the experimental trend.
The underlying dissipation mechanisms responsible for this velocity reduction are discussed in the next section.

\subsubsection{Impact force propagation and dissipation}

\begin{figure*}[htbp]
\begin{center}
\includegraphics[width=0.9\linewidth]{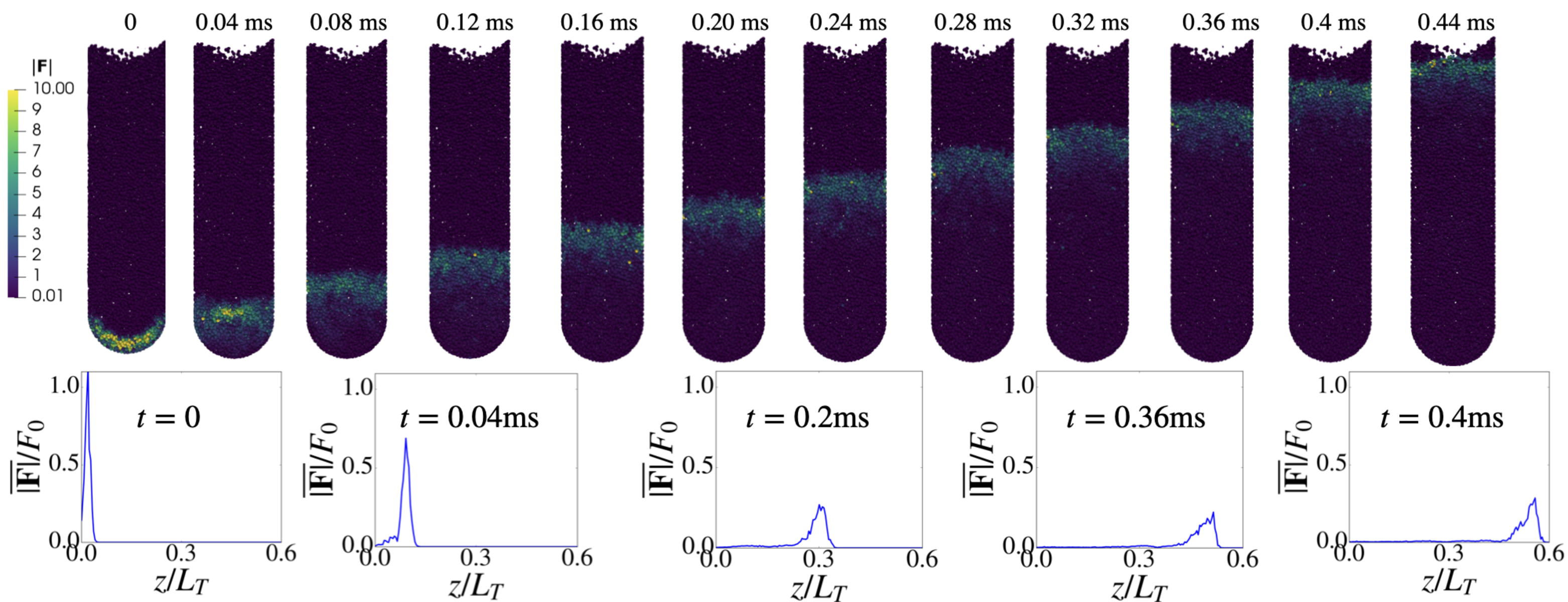}
\end{center}
\caption{\label{fig:7} (Color online) Time evolution of force propagation.}
\end{figure*}

To understand the transmission and dissipation of impulsive energy in the granular system, we analyzed the time evolution of force magnitude $|\mathbf{F}|$ on each particle and plane-averaged force $|\overline{\mathbf{F}}|$ on the $x$--$y$ plane.
As shown in Fig.~\ref{fig:7}, the peak of the impact force propagates upward and reduces as it approaches the meniscus.
This attenuation is due to inelastic particle--particle interactions.
When the granular pile is deeper (i.e., larger $L_{\rm G}$), the number of contact events increases, leading to higher energy dissipation.
This progressive loss of impact energy contrasts sharply with the response in low-viscosity liquids, where energy is efficiently focused toward the free surface \cite{Antkowiak2007, Kiyama2016}.

\begin{figure}[htbp]
\begin{center}
\includegraphics[width=0.9\linewidth]{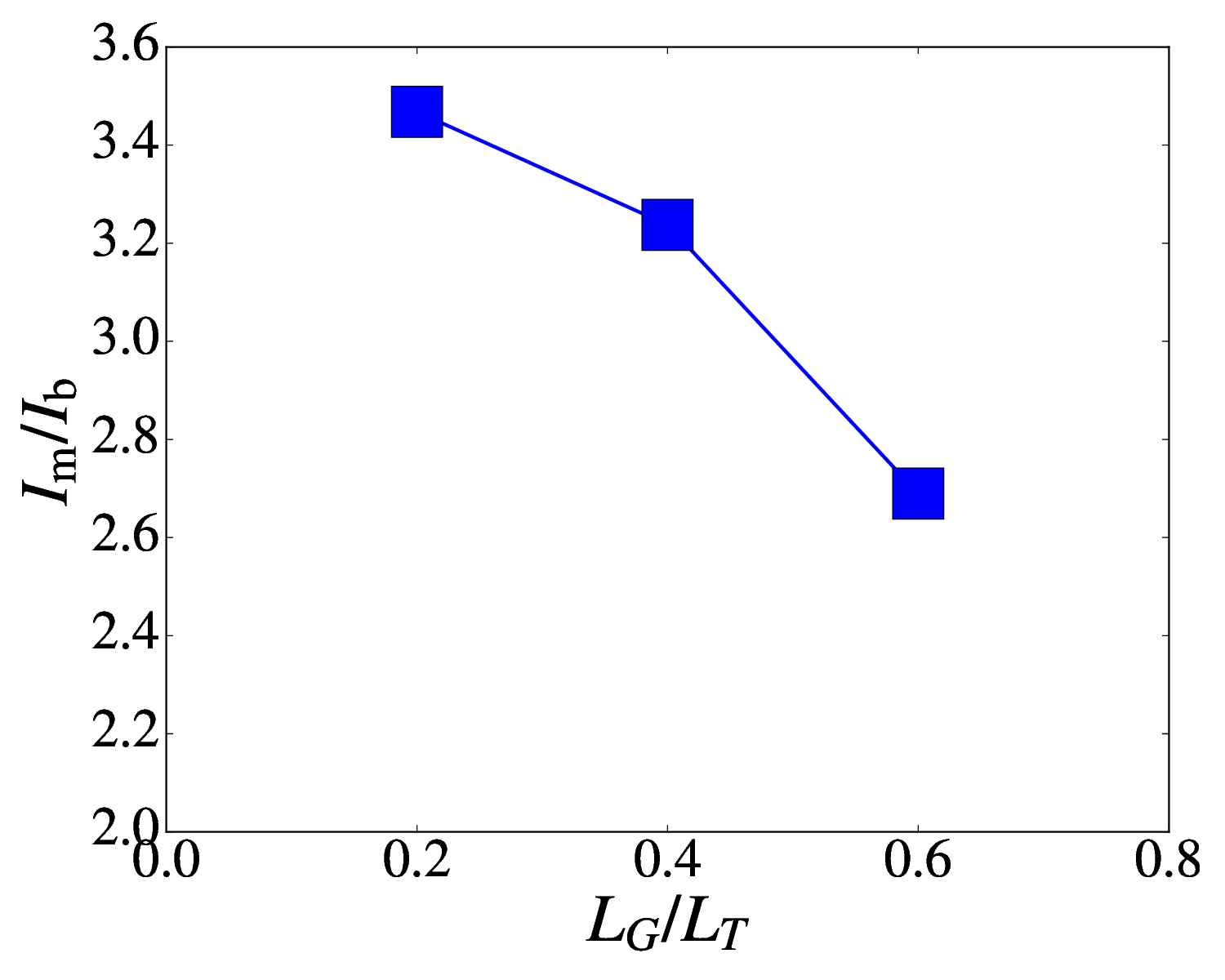}
\end{center}
\caption{\label{fig:8} Impulse $I_{\rm{m}}$ at meniscus scaled by impulse $I_{\rm{b}}$ at the bottom of tube according to granular pile height $z/L_{\rm{T}}$.}\end{figure}

To quantify this dissipation, we calculated the impulse transmitted through the granular column.
The impulse was defined as $I := \int_{0}^{t_{\rm jet}} |\overline{\mathbf{F}}| dt$, where $t_{\rm jet}$ is the instant when the stress wave reaches the meniscus.
Figure~\ref{fig:8} shows the ratio of the impulse at the meniscus, $I_{\rm m}$, to that at the base of the tube, $I_{\rm b}$.
As $L_{\rm G}$ increases, $I_{\rm m} / I_{\rm b}$ decreases gradually, being consistent with the observed reduction in jet velocity (Fig.~\ref{fig:3}).
This result demonstrates that in granular systems, the increasing dissipation with depth limits the transmission of momentum to the surface and thus the strength of the resulting jet.
This depth-dependent decay in impulse and velocity has not been reported in low-viscosity liquid jets, suggesting a dissipation mechanism that is unique to granular materials.

To further contextualize these findings, we analyze dissipation in both liquid and granular jets.
In liquid systems, the velocity field near the interface immediately after impact is largely unaffected by viscosity \cite{Onuki2018}, indicating that viscous effects have a negligible influence during early kinematic focusing.
However, once the interface begins to deform, jet formation becomes sensitive to liquid viscosity. In low-viscosity liquids, the jet velocity can reach nearly twice the impact velocity, $U_0$, due to efficient focusing, while in high-viscosity liquids, the velocity enhancement is almost suppressed.
In our experiments using low-viscosity silicone oil (10~mm$^2$/s), focusing-driven amplification was clearly observed.

Granular jets exhibit a fundamentally different dissipation behavior.
During the bulk transmission of the impulsive force, granular media show a large energy loss due to frequent interparticle contacts, especially in deeper layers.
Bulk dissipation limits the impulse that reaches the free surface, thereby limiting the jet velocity.
However, once the surface interface begins to deform and the jet emerges, the number of particle--particle contacts in the jet core rapidly decreases as the grains enter ballistic trajectories.
Hence, unlike high-viscosity liquids where dissipation continues during jet formation, granular jets experience a negligible energy loss during upward motion of the jet.
This distinction highlights a key difference between the two systems. While viscosity in liquids mainly affects jetting, dissipation in granular media primarily occurs during force propagation before jet formation.
Understanding this temporal separation of dissipation mechanisms is essential for interpreting jet dynamics across these two types of materials.

\section{\label{Conclusions}Conclusion}

We investigated the formation of focused granular jets induced by impulsive forces and compared them with impact-induced liquid jets through experiments and DEM simulations.
For a low granular pile height $L_{\rm G}$, the granular jet velocity closely matches that of liquid jets under equivalent conditions, indicating a similar underlying mechanism driven by kinematic focusing from a concave surface shape.
However, the jet behavior diverges drastically as $L_{\rm G}$ increases. Liquid jets show increased velocity due to cavitation at high $L_{\rm G}$, whereas granular jets exhibit reduced velocity with increasing pile height.
This contrasting behavior originates from different energy dissipation mechanisms.
In granular media, particle--particle contact leads to major dissipation during the transmission of impact forces, which is amplified in thicker layers.
This trend was confirmed through simulations that showed progressive attenuation of force propagation throughout the granular bed.
Furthermore, granular systems lack cavitation-like mechanisms that enhance jet velocity in liquids.
These findings highlight the distinct physical nature of granular jets and emphasize the role of internal dissipation in limiting jet formation in granular systems.

In future work, we will investigate the influence of the packing fraction and particle size on granular jet formation because these parameters directly affect compressibility and dissipation characteristics.
Moreover, recent studies on granular bubbles and droplet-like behavior in mixed particle systems \cite{Metzger2022, Guo2021} suggest potential analogies to cavitation in liquids.
Exploring such phenomena may provide deeper insights into the enhancement of jet velocity owing to gas-mediated or structural instabilities in granular beds.
Understanding these mechanisms will be essential for advancing the control and application of impulsively generated granular jets.

\begin{acknowledgments}
We thank Y. Sato, K. Masuda and K. Saito for helping with our experiments.
K.U.K. acknowledges support from Grant-in-Aid for Young Scientists (24K17024) from the Japan Society of the Promotion of Science (JSPS) and the Ihara Science Nakano Memorial Foundation.
Y.T. acknowledges support from a Grant-in-Aid for Scientific Research (A) (24H00289) from the Japan Society of the Promotion of Science (JSPS).
\end{acknowledgments}

%

\end{document}